\documentclass[journal]{IEEEtran}
\usepackage[T1]{fontenc}
\usepackage{cite}
\usepackage{amsmath,amssymb,amsfonts}
\usepackage{graphicx}
\usepackage{textcomp}
\usepackage{xcolor}
\usepackage{colortbl}
\usepackage{booktabs}
\usepackage{array}
\usepackage{multirow}
\usepackage{url}
\usepackage[hidelinks]{hyperref}
\usepackage{xspace}
\definecolor{cBlue}{HTML}{2457A6}
\definecolor{cTeal}{HTML}{15867B}
\definecolor{cAmber}{HTML}{D98A1D}
\definecolor{cRed}{HTML}{B63A2B}

\newcommand{\us}{\,\ensuremath{\mu}s\xspace}
\newcommand{\MHz}{\,MHz\xspace}
\newcommand{\dB}{\,dB\xspace}
\newcommand{\Hhat}{\widehat{H}}

\begin{document}

\title{FPGA-Based Wi-Fi Channel Estimation with a Reliability-Gated Binary Residual Network}

\author{Jixing~Zhou and Xinming~Huang%
\thanks{The authors are with the Department of Electrical and Computer Engineering,
Worcester Polytechnic Institute, Worcester, MA 01609 USA
(e-mail: jzhou1@wpi.edu; xhuang@wpi.edu).}%
\thanks{This work was supported by The MathWorks, Inc.}}

\markboth{}{}
\maketitle

\begin{abstract}
\boldmath
Wi-Fi receivers rely on accurate channel estimates to equalize received symbols and reduce the packet error rate (PER).
A learned estimator must deliver these estimates within microseconds while fitting within the receiver's logic budget.
We present RG-BRN, a reliability-gated binary residual network whose 14-convolution body takes no signal-to-noise-ratio (SNR) input while a closed-form SNR proxy gates its correction, and a channel-first field-programmable gate array (FPGA) engine storing each position's feature channels in one word.
On 802.11n Model B, RG-BRN reaches $-22.57$\dB{} normalized mean-square error (NMSE) power-averaged over \mbox{0--40\dB{}}, 6.2\dB{} below the openwifi smoother and within 0.2\dB{} of covariance-fitted linear minimum mean-square-error (LMMSE) estimation. Where the openwifi receiver estimator's simulated PER is 10--90\%, RG-BRN cuts its packet errors by 20.2\%.
\mbox{RG-BRN's} 1-bit-body design leads an identically trained floating-point twin by 2.2\dB{} NMSE under a $60^\circ$ inter-symbol phase offset.
One weight set trained on Models A--F leads pooled-covariance LMMSE in power-averaged NMSE on each (1.3\dB{} overall).
In register-transfer-level (RTL) simulation, the engine returns an isolated estimate in 7.09\us{} and meets the receiver's readiness bound with an 8.59-$\mu$s margin.
In a complete baseband-replay ZCU102 receiver, 7,680 read-back channel-memory words bit-match RTL simulation, the 1,600-packet PER lies inside software 95\% intervals, and full-design programmable-logic power averages 2.67\,W.
The project is available on GitHub at \url{https://github.com/JixingZhou2/rgbrn-ce-fpga}.

\end{abstract}


\begin{figure*}[!t]
\centering
\includegraphics[width=\textwidth]{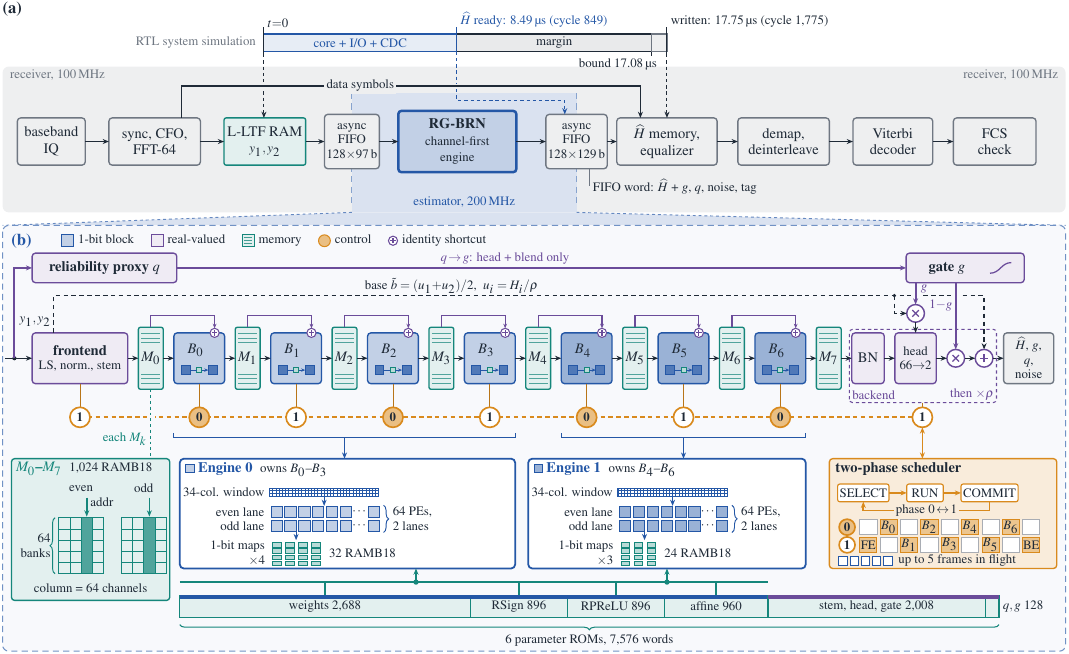}
\caption{Overall architecture. (a)~The 200-MHz estimator in the 100-MHz 802.11 receive chain, joined by asynchronous clock-crossing buffers; receive-cycle timing: ready at 8.49\us, readiness bound 17.08\us, write-back done at 17.75\us{} (first HT-data use at about 23.3\us). (b)~RG-BRN mapped onto the channel-first engine, the reliability path (proxy $q$, gate $g$) above the stage row.}
\label{fig:overview}
\end{figure*}

\section{Introduction}
\label{sec:intro}
\IEEEPARstart{C}{hannel} estimates govern the packet error rate (PER) of an IEEE 802.11 link and are due within microseconds: an orthogonal frequency-division multiplexing (OFDM) frame with a legacy preamble is first equalized with an estimate built from the two 3.2-$\mu$s symbols of the legacy long training field (L-LTF)~\cite{ieee2025std80211}.
A normalized mean-square error (NMSE) lowers the signal-to-noise ratio (SNR) to an effective ${\mathrm{SNR}/(1+\mathrm{NMSE}\cdot\mathrm{SNR})}$; at 10-dB packet SNR (11.1\dB{} on the active tones) on IEEE 802.11 task-group-n (TGn) Model~B (TGn-B)~\cite{erceg2004tgn} this loss is, to first order, 1.77\dB{} for the least-squares (LS) estimate and 0.72\dB{} for the openwifi receiver's L-LTF smoother~\cite{jiao2020openwifi,shi2022openofdm,openwifihw}.
Linear minimum mean-square-error (LMMSE) estimation~\cite{edfors1998svd} needs second-order statistics of the deployment channel. A refined estimate also has a deadline: in register-transfer-level (RTL) system simulation of the openwifi receiver, the equalizer first applies it to a high-throughput (HT) data symbol about 23.3\us{} after the first L-LTF read (signal fields use early LS), and our clock-crossing adapter enforces a readiness bound at 17.08\us{}.

Learned channel estimation~\cite{ye2018power,guler2025adafortitran} has reached field-programmable gate arrays (FPGAs) with multi-bit datapaths~\cite{haq2023lsdnn,haq2024lclsdnn,sharma2024lsidnn}, and 1-bit networks~\cite{hubara2016bnn,umuroglu2017finn} serve wireless physical-layer tasks~\cite{lu2021bcsinet,lu2022acrnet,shankar2024binarized}.
To our knowledge, within the surveyed literature, no prior work joins a deep estimator with 1-bit convolutions serving 0--40\dB{} SNR with one weight set, a dataflow that meets the deadline, and bit-exact board evidence in a complete receiver (Sec.~\ref{sec:related}).

RG-BRN, a reliability-gated binary residual network, and a channel-first engine provide these properties (Fig.~\ref{fig:overview}).
A closed-form reliability proxy $q$ from the two L-LTF symbols, a repeated-preamble SNR statistic~\cite{boumard2003snr,zivkovic2009preamble} insensitive to their phase offset, sets a gate $g$: the output is the LS average plus $(1-g)$ times the network's correction. The 14 binary convolutions receive no SNR input, and proxy and gate run in integer hardware on 1.7\% of the engine's synthesized look-up tables (LUTs).
Channel-first means all channels of one position in one memory word: one access returns all 64 channels of a fast Fourier transform (FFT) bin, and two block engines run the seven blocks under a two-phase schedule.

\emph{Accuracy without an SNR input.} On TGn-B, RG-BRN reaches a macro NMSE of $-22.57$\dB{} (error power averaged over 0--40\dB): 6.2\dB{} below the openwifi smoother, 10.7\dB{} below LS and within 0.19\dB{} of an LMMSE estimator fitted to the training covariance with the noise variance estimated from the two symbols (RG-BRN's NMSE is 0.18\dB{} lower at 0\dB). In identical single-stage training, the 1-bit-body network leads its 32-bit floating-point (FP32) twin by 2.2\dB{} under a $60^\circ$ common phase on the second L-LTF symbol, a residual carrier-frequency-offset (CFO) proxy, and is level with it zero-shot (TGn-B training only) on TGn A--F, trading 0.51\dB{} of TGn-B macro NMSE for 11.6$\times$ smaller packed parameters. In the same retrainings, the gate gains 5.1\dB{} over an ungated network at 40\dB{} on TGn-B and 1.9\dB{} in zero-shot A--F macro NMSE, and the 1-bit-body network leads real-valued dense networks on the LS average by at least 5.5\dB{} zero-shot on A--F.
In simulation, where the openwifi receiver estimator's PER is 10--90\% (modulation and coding scheme (MCS) 3 at 9--11\dB, MCS7 at 20 and 22\dB), RG-BRN removes 20.2\% of the packet errors of that estimator, which uses one HT training symbol for HT data, and 38.3\% of LS's: 387 errors in 1,000 packets against 485 and 627 (13.9\% and 29.9\% over the eight board-run points).
One weight set trained on a TGn A--F mixture leads a pooled-covariance LMMSE estimator in macro NMSE on every profile (A--F macro $-19.97$ against $-18.62$\dB). With TGn-B as the last of its three training stages, RG-BRN also leads a TGn-B-fitted LMMSE by 5.9\dB{} in A--F macro. On the board, a timing-closed build with the mixture weights makes 13.5\% fewer packet errors than the TGn-B build over 4,800 packets at 24 A--F transition points.

\emph{A channel-first engine.} The engine uses 256 processing-element (PE) datapaths and 1,080 18-kb block RAMs (RAMB18).
In RTL simulation, it returns an isolated estimate in 7.09\us{} (1,417 cycles at 200\MHz) and produces one estimate every 673 cycles under saturation.
In the complete receiver simulation, the estimate is ready 8.59\us{} before the readiness bound.

\emph{Board-verified integration.} On a ZCU102 board, the complete receiver processes replayed baseband: 7,680 channel state information (CSI) words read back from its channel memory (3,120 of them refined RG-BRN estimates) equal the RTL receiver's, the 1,600-packet board PER lies inside the software 95\% intervals at 8 of 8 points, and INA226 monitors measure a mean of 2.67\,W on the programmable-logic (PL) rails of the complete design.

\section{Related Work}
\label{sec:related}
For IEEE 802.11a, compressed multilayer perceptrons mapping the L-LTF to 52 channel coefficients run on a software-radio FPGA~\cite{soltani2022receiver}; LS-augmented and quantized multi-bit networks target Zynq platforms~\cite{haq2024lclsdnn,sharma2024lsidnn,brennsteiner2022receiver,lyu2026swiftchannel}.
Learned estimators in this journal use a RISC-V core, a vision-transformer processor and a 5G millimeter-wave FPGA design~\cite{guo2025riscvce,yun2026vitofdm,chundi2022fpgace}; low-rank LMMSE~\cite{edfors1998svd} is the classical reference.
Per-environment parameter sets for a fixed architecture are stored~\cite{haq2023lsdnn} or trained for reload~\cite{sharma2024lsidnn}; here one mixture-trained weight set leads a pooled-covariance LMMSE in macro NMSE on every TGn profile A--F~\cite{erceg2004tgn} (Sec.~\ref{sec:af}).

SNR conditioning is established: a noise-level subnet drives an LS-residual denoiser~\cite{li2023ndrnet}, a known SNR joins the features~\cite{guler2025adafortitran} or weights past slots~\cite{elazizi2026smartce}, and our earlier SNR-gated late-fusion shortcut for a diffusion estimator takes the SNR as an input~\cite{zhou2026gsldiffusion}.
Per-SNR-range networks~\cite{vanlier2020optcomnet} hold several parameter sets, an FPGA emitter identifier switches to binary weights at high estimated SNR~\cite{gan2022sasseinet}, SNR mismatch costs a model-driven receiver about 3\dB{} of mean-square error at 5\dB~\cite{gao2018comnet}, and the two-symbol noise statistic is standard~\cite{boumard2003snr,zivkovic2009preamble,wallace2012noise}.
Here the gate comes from the packet's own L-LTF symbols and one 61.34-KiB weight set spans 0--40\dB.

Binary image restoration with a real-valued head and tail~\cite{xia2023bbcu} is the dense-regression precedent for RG-BRN's 14 binary convolutions.
FINN streams binary networks through one engine per layer~\cite{umuroglu2017finn}, whereas RG-BRN time-shares two engines. Sub-8-bit FPGA hardware for learned wireless functions includes a fully unrolled ternary \mbox{1-D} modulation classifier~\cite{tridgell2019amc} and a 2-bit equalizer~\cite{defreitas2026fpgaeq}; 1-bit activations appear in a spiking receiver~\cite{gupta2025neuromorphicrx}.
A temporal-convolution accelerator puts consecutive samples in adjacent RAMB18 banks~\cite{carreras2020tcn} (see also~\cite{wu2024dilatedfpga}); the channel-first layout (channel-innermost, as in \cite{zhou2021implicit}) gives each bank a disjoint channel subset, so one column read serves all eight dilations.
Multi-CLP synchronizes engines with disjoint layer subsets at epoch boundaries~\cite{shen2017multiclp}, a dual-core overlay interleaves the layers of two images with ping-pong buffers per core~\cite{zhao2021dualopu}, and in-place per-layer buffers avoid ping-pong copies in a dilated temporal-convolution chip~\cite{denblanken2025chameleon}; here a global fence alternates the parity of the blocks in flight on two shared engines, so no ping-pong copy exists.
Board PER has been reported for a classical estimator in an 802.11 FPGA receiver~\cite{havinga2023hls}, and a binary/ternary-weight accelerator in this journal measures dynamic power on a ZCU102~\cite{hu2026bapetransformer}.
For a learned estimator, this work reports bit-exact board read-back against the RTL, a 1,600-packet board PER over replayed baseband and INA226-measured power~\cite{ti_ina226}.
To our knowledge, within the surveyed literature, no prior work derives a closed-form, phase-invariant SNR proxy from the two L-LTF symbols of the same packet and uses it, through a sigmoid gate that enters a 1-bit-body network only at its head input and output blend, to weight a learned residual correction of the LS-average estimate, with proxy and gate realized in integer hardware; nor does any join channel-first banked memory, two shared engines owning disjoint residual blocks and a fenced two-phase cross-frame schedule without ping-pong copies.

\section{RG-BRN: Reliability-Gated Binary Residual Network}
\label{sec:method}

\subsection{Observations and the reliability proxy}
Removing the $\pm1$ training sign from the FFT outputs $y_1,y_2\in\mathbb{C}^{64}$ of the two L-LTF symbols gives the LS observations $H_1,H_2$; $\mathcal{A}$ denotes the 52 active bins. With $\rho$ the root-mean-square value of their average over $\mathcal{A}$ and $u_i=H_i/\rho$, the reliability proxy and gate are
\begin{align}
P&=\sum_{k\in\mathcal{A}}\!\big(|u_1[k]|^2+|u_2[k]|^2\big),\;\; C=\sum_{k\in\mathcal{A}}u_1[k]\,u_2^{*}[k],\label{eq:pc}\\
q&=10\log_{10}\frac{2|C|}{P-2|C|},\qquad g=\sigma\!\Big(\frac{q-\mu}{s}\Big),\label{eq:gate}
\end{align}
where $\sigma$ is the logistic function, $\mu$ and $s=\mathrm{softplus}(s_{\mathrm{raw}})+0.1$ are learned, and the hardware clips $q$ to $[-30,60]$\dB.
We call $q$ a reliability proxy because its gate weights the LS observation (cf.\ a subcarrier reliability test~\cite{gizzini2020trfi}).
As $P-2|C|$ is the noise left after phase alignment and $2|C|$ the aligned signal, $q$ is a repeated-preamble statistic~\cite{boumard2003snr,zivkovic2009preamble} that tracks the SNR on the active tones (on average 1.0--1.1\dB{} above the SNR of Sec.~\ref{sec:protocol}); up to 25\dB{} its mean bias stays within 0.12\dB{} (0.6--1.1\dB{} root-mean-square error) on the TGn-B test set and within 0.3\dB{} on every TGn profile A--F.
Because $|C|$ discards a constant inter-symbol phase, this phase-consistent $q$ is invariant to the receiver gain and to the common phase that a residual CFO adds. Rotating the second L-LTF symbol by 5 to 60$^\circ$ (4.3--52.1\,kHz CFO-equivalent) changes $q$ by at most 0.015\dB{} per packet, whereas the L2 proxy, symbol-average over symbol-difference power, drifts by 19.7\dB{} on average at 60$^\circ$; there the deployed RG-BRN keeps a \mbox{10.62-dB} macro gain over LS (10.66\dB{} at 0$^\circ$; Sec.~\ref{sec:acc}).
In integer hardware (Sec.~\ref{sec:relhw}), $q$ and $g$ take 207 cycles after the last accumulation.

\subsection{Gated residual output}
\label{sec:gate}
The proxy enters only at the head input and the output blend, so the 14 binary convolutions take no SNR input; with the normalized LS average $\tilde b=\tfrac12(u_1+u_2)$,
\begin{equation}
\Hhat=\rho\,\big[\,\tilde b+(1-g)\,\tilde\Delta\,\big],\qquad \tilde\Delta=\mathrm{Head}\big([F,\;g\,\tilde b]\big),
\label{eq:out}
\end{equation}
with $F$ the 64 body feature channels and two head input channels $g\tilde b$. On the TGn-B test set the gate $g$ rises from 0.27 at 0\dB{} to 0.93 at 40\dB{} while the applied correction falls from 0.37 to $3.7\times10^{-5}$ of the channel power.

\subsection{Binary residual body}
\label{sec:body}
\begin{figure*}[!t]
\centering
\includegraphics[width=\textwidth]{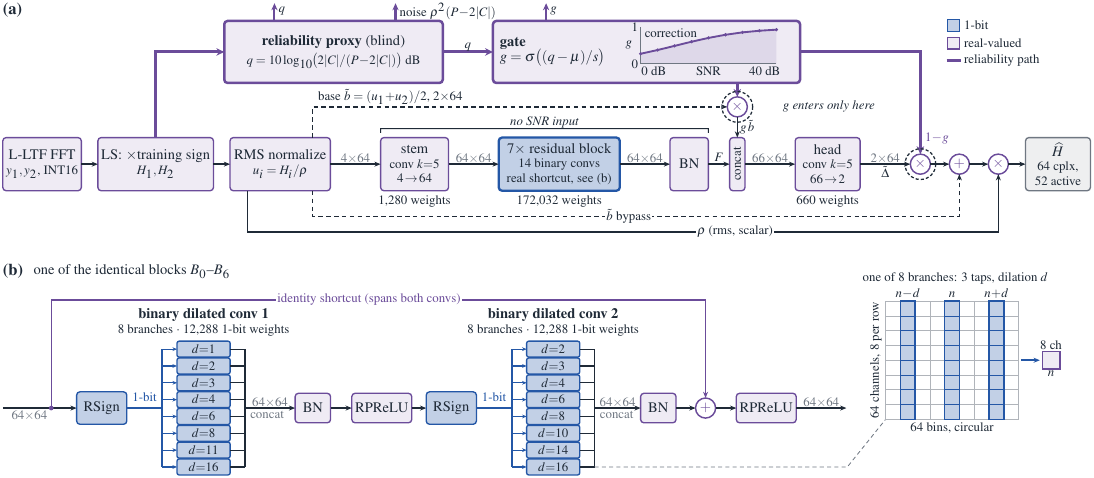}
\caption{RG-BRN. (a)~Data path; the reliability path (heavy purple: proxy $q$, gate $g$ with its curve over SNR) enters only the head input and the $(1-g)$ blend (dashed rings). (b)~One residual block; one branch on the channel-by-bin grid. The 14 binary convolutions (blue) receive no SNR input.}
\label{fig:network}
\end{figure*}

The network (Fig.~\ref{fig:network}) chains a real-valued 5-tap stem (4$\to$64), seven residual blocks, batch normalization (BN) and a 5-tap head (66$\to$2).
A block applies RSign (a learned-threshold sign), a binary convolution and BN twice, with RPReLU (parametric ReLU with learned shifts)~\cite{liu2020reactnet} between the two and after the sum with a real-valued identity shortcut~\cite{liu2018bireal} spanning both; each convolution has eight parallel circular 3-tap branches reading all 64 channels.
It has 179,544 parameters, 172,032 of them 1-bit (95.8\%); its packed inference form is 62,816\,B (61.34\,KiB), 11.6$\times$ below the FP32 packing, and its convolutions need 11,010,048 binary products and 124,160 real multiply-accumulates per estimate.

\subsection{Training and integer export}
\label{sec:train}
A three-stage curriculum with the straight-through estimator gives the deployed weights: a synthetic-channel mixture with delays up to 0.75\us{} (80 epochs), then its receiver-domain version and TGn-B (40 epochs each).
The last two stages minimize $\mathrm{NMSE}+0.1\,L_{\mathrm{eq}}+0.05\,L_{\mathrm{exc}}$, where \mbox{$L_{\mathrm{eq}}=\mathrm{E}\big[|H|^2|e|^2/(|H|^2{+}0.05\bar p)^2\big]$} is an equalization-sensitivity proxy on the reference channel $H$ (error $e$; $\bar p$: mean of $|H|^2$ over $\mathcal A$; $\mathrm{E}$: mean over active bins and training pairs) and $L_{\mathrm{exc}}$ penalizes, at training SNRs of 25\dB{} and higher, corrections above twice the LS error.
The TGn-B pairs are receiver-domain: an approximate 16-bit integer model of the receiver front end takes timing and CFO decisions from the noisy observation, and the label is the noiseless L-LTF average at the same decisions (54,000 training, 6,912 validation and 13,824 test pairs at 0 to 40\dB{} in 5-dB steps).
The final-stage checkpoint is the deployed model, exported to an integer model that the RTL reproduces bit for bit (Sec.~\ref{sec:verify}).

\section{Channel-First Engine}
\label{sec:arch}

\subsection{Organization}
Two shared block engines execute the seven blocks between the frontend and the backend (Fig.~\ref{fig:overview}b).
Eight 24-bit block-boundary maps (64 channels $\times$ 64 bins) and seven 1-bit maps sit in block RAM, one read port per bank: the 64-way column bandwidth, not capacity, sets the 1,080 RAMB18.
Engine~0 owns $B_0$--$B_3$ and engine~1 owns $B_4$--$B_6$; an engine runs one convolution at a time on 64 channel-parallel PEs of two spatial lanes each (even and odd FFT bins), 128 datapaths and 24,576 binary products per cycle. The frontend normalizes the input while 64 stem lanes write map~0; the backend streams map~7 into the parallel head, blend and rescaling.

\begin{figure*}[!t]
\centering
\includegraphics[width=\textwidth]{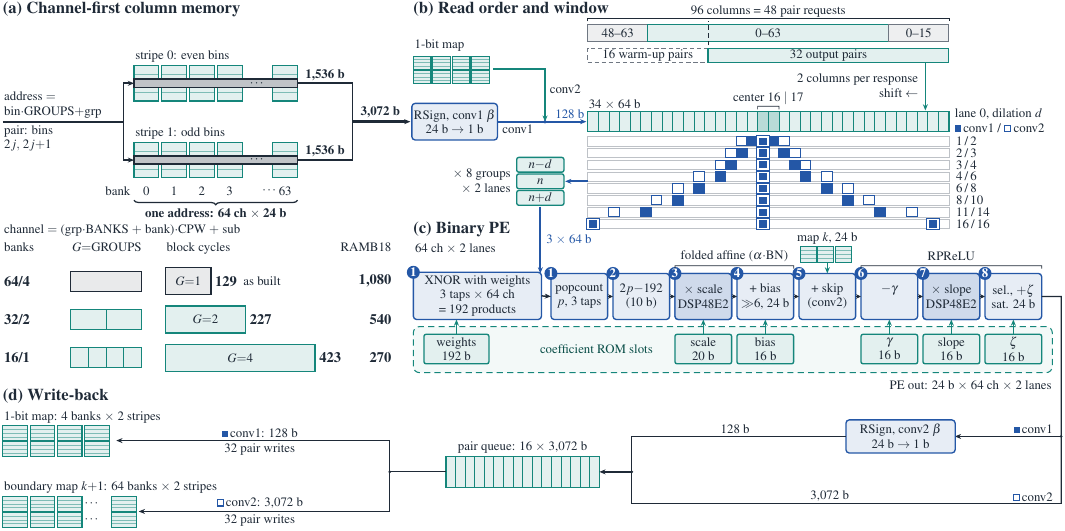}
\caption{Datapath of one engine. (a)~Column memory: one address reads a 64-channel column from each of the even and the odd stripe; bank profiles trade memory for block cycles. (b)~Read order and 34-column window with dilated taps. (c)~Binary PE: eight stages ($\beta$: threshold; $\alpha$: weight scale; $\gamma$, $\zeta$: shifts). (d)~Write-back through the pair queue.}
\label{fig:dataflow}
\end{figure*}

\subsection{Column-per-access memory}
Channel-first memory feeds the 64 PEs one column per address: all channels of one position share one memory word (bin-major, channel index innermost), 1,536 bits for a boundary map and 64 bits for a binary map (Fig.~\ref{fig:dataflow}a).
With \textsc{banks} banks per map, \textsc{cpw} channels per physical word (1 for boundary maps, 16 for binary maps) and $\textsc{groups}=64/(\textsc{banks}\cdot\textsc{cpw})$, \eqref{eq:addr} maps channel $c$ and bin $n$:
\begin{equation}
\begin{aligned}
c&=(\mathrm{grp}\cdot\textsc{banks}+\mathrm{bank})\cdot\textsc{cpw}+\mathrm{sub},\\
\mathrm{addr}&=n\cdot\textsc{groups}+\mathrm{grp}.
\end{aligned}
\label{eq:addr}
\end{equation}
Banks hold disjoint channels (no replicas); each map has an even and an odd stripe, so a pair transaction returns bins $(2j,2j{+}1)$.
The default 64/4 profile (64 boundary, 4 binary banks) reads a column per cycle; 32/2 and 16/1 take 2 and 4 cycles per column and halve or quarter the map memory (540 and 270 RAMB18, at 2,105 and 3,547 cycles; Sec.~\ref{sec:dse}).

\subsection{Window and binary PE}
A convolution reads 96 columns (48 pair requests) in the order $48..63\,|\,0..63\,|\,0..15$ (Fig.~\ref{fig:dataflow}b), so the circular padding is the read order itself, and tap registers at $-d$, $0$ and $+d$ serve all eight dilations from one column read of a $34\times64$-bit window.
Each lane of a PE consumes three 64-bit tap vectors and a 192-bit weight word per cycle and finishes in eight stages (Fig.~\ref{fig:dataflow}c); the scale and slope stages each use a digital signal processing (DSP) slice of type DSP48E2.
For tap vectors $a_\tau$ and weight vectors $w_\tau$, $\tau\in\{-d,0,d\}$, one issue computes
\begin{equation}
\begin{aligned}
p&={\textstyle\sum_{\tau}}\,\mathrm{popc}\big(\overline{a_\tau\oplus w_\tau}\big),\quad x=2p-192,\\
z&=\lfloor(\lambda x+2^{6}\theta+2^{5})/2^{6}\rfloor,\quad v=\mathrm{sat}\big(\psi(m-\gamma)+\zeta\big),
\end{aligned}
\label{eq:pe}
\end{equation}
with $m=z$ in the first convolution and $m=\mathrm{sat}(z+r)$ ($r$: shortcut from the block input) in the second; $\psi(t)=t$ for $t>0$ and $\lfloor(\kappa t+2^{10})/2^{11}\rfloor$ otherwise, and $\mathrm{sat}$ saturates to the 24-b activation format (10 fraction bits).
The next convolution takes the bit $v\ge\beta$.
Per channel, the multiplier $\lambda$ (20~b, 16 fraction bits) and bias $\theta$ fold the weight scale $\alpha$ and BN; $\theta$, $\gamma$, $\zeta$ and $\beta$ are 16~b with 10 fraction bits, and the negative slope $\kappa$ is 16~b with 11.
Stem and head weights are 16~b (12 fraction bits); all weights and constants are ROMs in logic.
Results return to the maps through a pair queue (Fig.~\ref{fig:dataflow}d).

\begin{figure}[!t]
\centering
\includegraphics[width=\columnwidth]{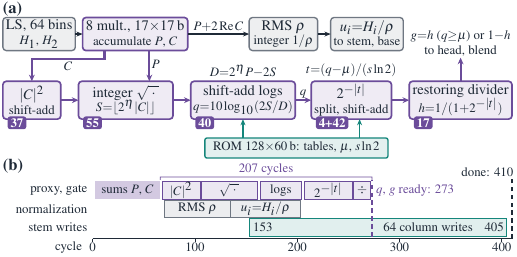}
\caption{Reliability path in integer hardware. (a)~Proxy and gate circuit beside the normalization (gray); boxed numbers: iteration cycles, 207 in total with control; $2^{\eta}$ normalizes $P$. (b)~Frontend timing from launch (stem writes and completion: RTL trace; other bars: state sequence): $q$ and $g$ are ready while the stem writes.}
\label{fig:relpath}
\end{figure}
\subsection{Reliability path in integer hardware}
\label{sec:relhw}
The frontend accumulates $P$ and $C$ of~\eqref{eq:pc} on the unnormalized observations, to which $q$ is invariant, in one pass over the 64 bins (Fig.~\ref{fig:relpath}); the same sums give the normalization energy $P+2\,\mathrm{Re}\,C$.
The proxy takes $|C|$ as the integer square root $S=\lfloor 2^{\eta}|C|\rfloor$, with $2^{\eta}$ normalizing $P$, so the phase of $C$ never reaches $q$.
Apart from eight accumulation multipliers and one constant multiplication, the proxy-gate circuit uses shift-and-add units and a $128\times60$-bit ROM, in all 3,192 LUTs (1.7\% of the default design's 185,093 synthesized LUTs), 1,448 flip-flops (FFs) and 11 DSP slices.
It runs in parallel with the normalization and stem, adding no latency, and on all 13,824 test pairs its bit-exact integer model keeps $q$ and $g$ within half an output step of double precision ($1/32$\dB{} and $2^{-16}$).

\subsection{Two-phase cross-frame schedule}
\label{sec:schedule}
\begin{figure*}[!t]
\centering
\includegraphics[width=\textwidth]{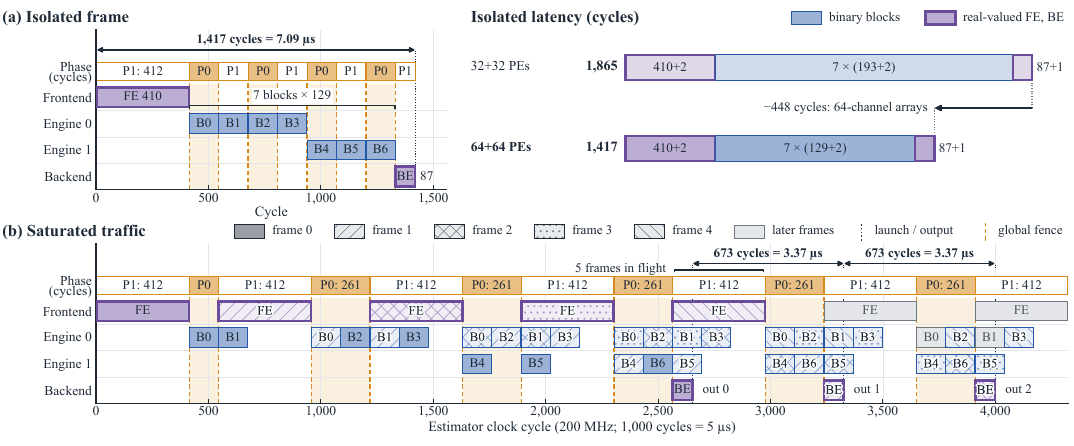}
\caption{Two-phase cross-frame schedule from RTL simulation traces (FE, BE: frontend, backend). (a)~One isolated frame (1,417 cycles); right: isolated latency for 32+32 and 64+64 PEs (64/4 banks), split into frontend, seven blocks and backend (hand-overs and output cycle included). (b)~Saturated traffic (16-frame RTL run with test-bench back-pressure on some outputs): outputs of frames 0--2 (shown) and 7--12 are 673 cycles apart (phase~1, P1: 412; phase~0, P0: 261).}
\label{fig:schedule}
\end{figure*}

The schedule keeps up to five frames in flight (holding a feature map; up to four jobs execute at once).
Blocks of one frame run in sequence (every output bin's receptive field is the whole ring), blocks of different frames in parallel.
The scheduler alternates two phases (Fig.~\ref{fig:schedule}): phase~0 runs $B_0,B_2,B_4,B_6$; phase~1 runs the frontend, $B_1,B_3,B_5$ and the backend.
Block $k$ reads map $k$ and writes map $k{+}1$ and a phase holds one parity of $k$, so no boundary map is both read and written in a phase, and two engines share the 15 maps without ping-pong copies.
A SELECT--RUN--COMMIT machine reserves a phase's eligible jobs, waits for all (global completion fence) and advances each map's ownership and 2,560-bit frame context by one slot.

In RTL simulation an isolated frame takes 1,417 cycles (7.09\us{} at 200\MHz) from frontend acceptance to output-valid: frontend 410, seven blocks of 129, eight hand-overs of 2, backend 87, one output cycle.
For the 64/4 bank profile, doubling the PE count per engine from 32 to 64 reduces each block from 193 to 129 cycles.
The saturated output interval is 673 cycles (3.37\us; phase~1: 412, phase~0: 261).

\subsection{Deadline-aware integration}
\label{sec:integration}
The engine runs at 200\MHz{} behind a clock-domain-crossing (CDC) adapter in the 100-MHz receiver (Fig.~\ref{fig:overview}a).
The adapter streams the 64 L-LTF beats through a 97-bit asynchronous first-in first-out (FIFO) buffer, writes an early LS estimate into the receiver's channel memory while the engine works, and accepts the refined estimate through a 128$\times$129-bit FIFO that also carries $q$, $g$, the noise estimate and a 32-bit job tag, checked on return.
In RTL simulation of the complete receiver (receive-domain cycles at the nominal clocks, not board time; cycle 0: first L-LTF read) the refined estimate is ready at cycle 849 (8.49\us): the 1,417 core cycles (708.5 receive cycles) plus about 141 cycles of input, output and CDC handshakes.
The adapter enforces readiness at cycle 1,708 (17.08\us) and raises a protocol error otherwise. The estimate is ready 859 cycles (8.59\us) before the bound.
After the bound, the adapter overwrites the early LS values (done at cycle 1,775, 17.75\us), and the equalizer, which never waits, applies the refined estimate from the first HT data symbol on (about 23.3\us).

\section{Results}
\label{sec:results}

\subsection{Protocol}
\label{sec:protocol}
\emph{Accuracy.} Every estimator is scored on the same 13,824 TGn-B test pairs of Sec.~\ref{sec:train} (1,536 per SNR point), disjoint from training and validation; Table~\ref{tab:acc} lists what each knows at run time.
The scope is single-antenna 20-MHz frames over static TGn-B channels without added CFO, phase noise or in-phase/quadrature (IQ) imbalance (SNR: time-domain packet power over injected noise variance). NMSE uses the front-end model's decisions on the captures with both FFT windows inside the received L-LTF (20.1\% at 0\dB, 99.8\% or more from 5\dB{}; the PER counts every packet).
The macro NMSE is the equal-weight mean of the nine per-SNR linear means of $\sum_{\mathcal A}|\Hhat{-}H|^2/\sum_{\mathcal A}|H|^2$, in dB, which weights the weakest links most.
Intervals are 95\% paired channel-bootstrap intervals over the test channels; network NMSE values are floating-point evaluations (integer RG-BRN: Sec.~\ref{sec:acc}).

\emph{Retrained controls.} Table~\ref{tab:abl} gives seed means of the network and variants retrained in one 80-epoch TGn-B stage with an active-tone mean-square-error loss.
The common-phase stress, a residual-CFO proxy, rotates the second L-LTF symbol by $\phi\le60^\circ$ (52.1\,kHz), with label $H(1{+}e^{j\phi})/2$; A--F scoring (zero-shot except on B) runs these checkpoints on 13,824 TGn A--F pairs (2,304 per profile; macro: mean of the six per-profile linear macros).

\emph{Packet error rate.} 128-byte frames at MCS3 and MCS7, 20\,MHz, static TGn-B channels, 200 packets per point (the same 200 channels and payloads at every SNR point of an MCS, independent noise).
The complete openwifi OpenOFDM receiver runs in RTL simulation with RG-BRN and LS as double-precision C++ models that also serve the signal fields (early LS on the board); intervals are Wilson 95\%~\cite{wilson1927probable}.
The openwifi receiver estimator uses the openwifi smoother for the signal fields and, for HT data, a 3-tap-smoothed estimate from the single HT long training field (HT-LTF) symbol; LS and RG-BRN use the two L-LTF symbols and the HT-LTF edge tones.
Pooled counts take the points where the openwifi receiver estimator's PER lies in $(0.1,0.9)$ (MCS3 at 9, 10, 11\dB; MCS7 at 20, 22\dB), with 95\% error-reduction intervals from a bootstrap over the 200 channels that both MCS share.
SNR gains at 50\% PER are interpolated linearly in PER, with paired channel-bootstrap intervals (both bootstraps: $10^6$ resamples).

\emph{Board runs.} On the ZCU102 (XCZU9EG, Vivado 2025.2), direct memory access (DMA) streams baseband IQ from the processing system's external memory to the RG-BRN receiver at 20\,Msps. The host reads back frame-check results and channel-memory words.

\subsection{Accuracy and ablation}
\label{sec:acc}
\begin{figure*}[!t]
\centering
\includegraphics[width=\textwidth]{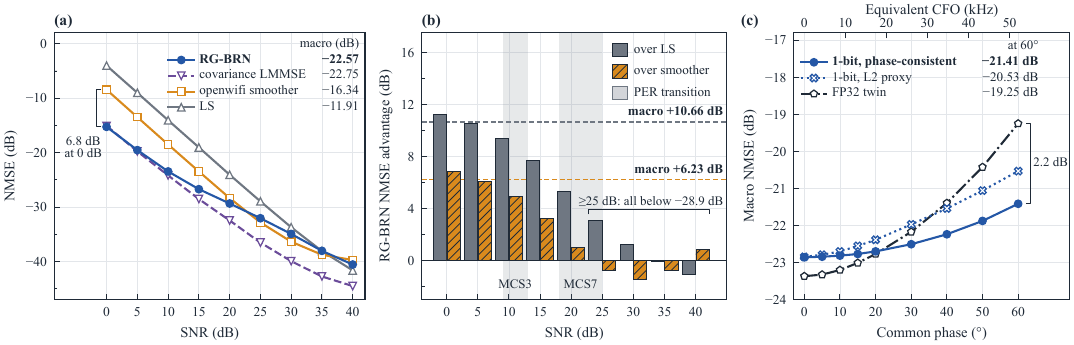}
\caption{NMSE on TGn-B (13,824 test pairs). (a)~NMSE versus SNR. (b)~Per-SNR NMSE advantage of RG-BRN over LS and the smoother (dashed: macro; bracket: NMSE of all three from 25\dB); gray: PER transitions of Fig.~\ref{fig:per} (MCS3 9--13\dB, MCS7 18--25\dB). (c)~Macro NMSE under a common phase on the second L-LTF symbol (top: equivalent CFO), single-stage retrainings (seeds: phase-consistent 3, L2 proxy 1, FP32 2).}
\label{fig:nmse}
\end{figure*}

At 10\dB, RG-BRN already reaches the NMSE that the openwifi smoother reaches at 15\dB{} and LS at 19.4\dB, and its active-tone effective-SNR cost there is 0.24\dB, against 1.77\dB{} (LS) and 0.72\dB{} (smoother).
Its macro NMSE is $-22.57$\dB{} (Fig.~\ref{fig:nmse}a), 6.23\dB{} below the smoother ($-16.34$\dB), 10.66\dB{} below LS ($-11.91$\dB) and within 0.19\dB{} (paired interval 0.10--0.26\dB) of a covariance-fitted LMMSE.
\begin{table*}[!t]
\centering
\caption{NMSE (\textnormal{dB}) on TG\textnormal{n}-B (13,824 Test Pairs)}
\label{tab:acc}
\scriptsize
\setlength{\tabcolsep}{1.25pt}
\providecommand{\rowcolor}[1]{}
\begin{tabular}{>{\raggedright\arraybackslash}p{2.1cm}>{\raggedright\arraybackslash}p{2.15cm}rrrrrrrrr>{\hspace{5pt}}r>{\hspace{3pt}}r>{\raggedleft\arraybackslash}p{2.5cm}}
\toprule
 & & \multicolumn{9}{c}{NMSE at SNR (dB)} & \multicolumn{2}{c}{Mean over SNR} & \\
\cmidrule(lr){3-11}\cmidrule(lr){12-13}
Estimator & Also known at run time & 0 & 5 & 10 & 15 & 20 & 25 & 30 & 35 & 40 & Macro & Band & Macro minus RG-BRN [95\% interval] \\
\midrule
LS (pair average) & -- & $-4.01$ & $-9.01$ & $-14.05$ & $-19.04$ & $-24.06$ & $-28.95$ & $-33.71$ & $-38.08$ & $-41.63$ & $-11.906$ & $-9.368$ & $+10.66$\,[$10.58$,{\kern1pt}$10.74$] \\
openwifi smoother & 3-tap average & $-8.44$ & $-13.48$ & $-18.50$ & $-23.46$ & $-28.35$ & $-32.83$ & $-36.36$ & $-38.76$ & $-39.73$ & $-16.336$ & $-13.804$ & $+6.23$\,[$6.15$,{\kern1pt}$6.31$] \\
Span-7 smoother & validated over 3/5/7 & $-11.72$ & $-16.69$ & $-21.36$ & $-25.52$ & $-28.54$ & $-30.01$ & $-30.47$ & $-30.89$ & $-30.61$ & $-19.383$ & $-16.983$ & $+3.18$\,[$3.11$,{\kern1pt}$3.24$] \\
DFT denoiser & validated 6 taps & $-12.86$ & $-17.50$ & $-21.26$ & $-23.79$ & $-25.07$ & $-25.37$ & $-25.40$ & $-25.52$ & $-25.17$ & $-19.724$ & $-17.733$ & $+2.84$\,[$2.79$,{\kern1pt}$2.89$] \\
PCA denoiser & basis, rank 4 & $-14.41$ & $-19.28$ & $-23.54$ & $-26.31$ & $-27.51$ & $-27.98$ & $-29.12$ & $-29.53$ & $-27.47$ & $-21.601$ & $-19.468$ & $+0.96$\,[$0.87$,{\kern1pt}$1.06$] \\
Learned linear map & map, validated ridge & $-14.83$ & $-19.67$ & $-23.91$ & $-26.83$ & $-28.15$ & $-28.78$ & $-29.76$ & $-30.18$ & $-28.32$ & $-22.049$ & $-19.886$ & $+0.52$\,[$0.43$,{\kern1pt}$0.61$] \\
Covariance LMMSE & covariance, noise est. & $-15.09$ & $-19.77$ & $-24.18$ & $-28.53$ & $-32.46$ & $-36.53$ & $-39.94$ & $-42.75$ & $-44.51$ & $-22.754$ & $-20.239$ & $-0.19$\,[$-0.26$,{\kern1pt}$-0.10$] \\
\midrule
\rowcolor{cBlue!10}
\textbf{RG-BRN} & deployed weights & $-15.27$ & $-19.54$ & $-23.45$ & $-26.72$ & $-29.34$ & $-32.05$ & $-34.94$ & $-38.04$ & $-40.56$ & $-22.566$ & $-20.118$ & -- \\
\bottomrule
\end{tabular}
\par\vspace{2pt}
\parbox{\linewidth}{\footnotesize Macro: equal-weight mean of the nine per-SNR linear means, in dB; Band: the same over 0--20\dB. All estimators take the two L-LTF symbols; column~2 lists what each also knows at run time (statistics fitted on the training split; validated: chosen on the validation split). Last column: estimator minus RG-BRN macro with its paired 95\% interval. 95\% bootstrap half-widths per SNR point: RG-BRN at most 0.22\dB, all rows at most 1.52\dB. DFT: discrete Fourier transform; PCA: principal-component analysis.\par}
\end{table*}
\unskip 
The gain over the smoother spans 0--20\dB, largest at 0\dB{} (6.83\dB; Fig.~\ref{fig:nmse}b); from 25\dB{} the gate shrinks the correction to a mean weight $1-g$ of 0.07--0.21.
In NMSE, RG-BRN is 0.18\dB{} below the covariance LMMSE at 0\dB{} and within 0.12\dB{} of it in the 0--20-dB band mean (Table~\ref{tab:acc}), with one fixed \mbox{61.34-KiB} parameter set and no covariance at run time.
Its macro NMSE is also below those of the four other validated or fitted estimators, by 0.52\dB{} (learned linear map) to 3.18\dB{} (\mbox{span-7} smoother; paired intervals exclude zero); in the 25--40-dB band mean, its NMSE is 6.1\dB{} below the linear map's.

The retrained controls isolate each design choice (Table~\ref{tab:abl}, Fig.~\ref{fig:nmse}c).
The gate lowers the high-SNR floor of the ungated network by 5.1\dB{} at 40\dB{} on TGn-B ($-40.08$ against $-34.96$\dB) and keeps a 1.9-dB macro-NMSE lead over that network zero-shot on A--F.
The phase-consistent proxy matches the L2 proxy at $0^\circ$ and leads it at $60^\circ$ by 0.88\dB{} in macro NMSE and by 13.0\dB{} at 40\dB{} ($-39.10$ against $-26.13$\dB).
With the deployed weights frozen, the blind $q$ is as good as the true packet SNR (macro NMSE within 0.014\dB) and tolerates a $\pm3$-dB offset (at most 0.24\dB); at $60^\circ$ it beats an L2-proxy gate by 1.9\dB{} ($-21.25$ against $-19.38$\dB).
The 1-bit-body network leads its FP32 twin by 2.2\dB{} at $60^\circ$ ($-21.41$ against $-19.25$\dB; ahead at every SNR point, by 1.3--4.2\dB) and is level with it zero-shot on A--F ($-15.76$ against $-15.64$\dB), trading 0.51\dB{} of \mbox{TGn-B} macro NMSE for 11.6$\times$ smaller packed parameters.
Real-valued dense networks on the LS average (one 52-unit or two 104-unit hidden layers, sized as in~\cite{haq2023lsdnn}), trained alike, trail the 1-bit-body network by 0.84 and 0.32\dB{} in \mbox{TGn-B} macro NMSE and by at least 9\dB{} at 40\dB{} and 5.5\dB{} zero-shot on A--F, where both fall behind LS: from 10\dB{} upward their A--F NMSE stays within 0.2\dB{} of $-10.4$\dB.
The deployed three-stage weights, trained with no TGn A or C--F data, reach $-17.58$\dB{} zero-shot on A--F, 1.8\dB{} ahead of the single-stage retraining.
Their integer model, which the RTL reproduces bit for bit, reaches $-22.55$\dB{} TGn-B macro NMSE, 0.014\dB{} from the floating-point evaluation (interval 0.002--0.026\dB).
%
\begin{table}[!t]
\centering
\caption{Ablation NMSE (\textnormal{dB}) on TG\textnormal{n}-B and TG\textnormal{n} A--F}
\label{tab:abl}
\scriptsize
\setlength{\tabcolsep}{3.9pt}
\providecommand{\rowcolor}[1]{}
\begin{tabular}{lccccr}
\toprule
 & \multicolumn{2}{c}{TG\textnormal{n}-B} & TG\textnormal{n}-B, $60^\circ$ & Zero-shot & \\
\cmidrule(lr){2-3}\cmidrule(lr){4-4}\cmidrule(lr){5-5}
Variant & Macro & 40\dB & Macro & A--F & KiB \\
\midrule
\multicolumn{6}{l}{\emph{Retrained: single stage, 80 epochs}} \\
\textbf{1-bit reference} ($n{=}3$) & $-22.86$ & $-40.08$ & $-21.41$ & $-15.76$ & 61.34 \\
FP32 twin ($n{=}2$) & $-23.37$ & $-42.28$ & $-19.25$ & $-15.64$ & 708.8 \\
No gate ($n{=}2$) & $-22.78$ & $-34.96$ & $-21.36$ & $-13.86$ & 61.34 \\
L2 proxy ($n{=}1$) & $-22.84$ & $-40.22$ & $-20.53$ & $-15.72$ & 61.34 \\
\midrule
\multicolumn{6}{l}{\emph{Real-valued baselines, same split and recipe}} \\
Dense, 1 layer ($n{=}3$) & $-22.02$ & $-31.05$ & $-20.67$ & $-10.23$ & 42.86 \\
Dense, 2 layers ($n{=}3$) & $-22.54$ & $-30.88$ & $-21.16$ & $-10.17$ & 128.0 \\
\midrule
\multicolumn{6}{l}{\emph{Deployed: three-stage curriculum, 80+40+40 epochs}} \\
\rowcolor{cBlue!10}
\textbf{RG-BRN deployed} & $-22.57$ & $-40.56$ & $-21.25$ & $-17.58$ & 61.34 \\
\bottomrule
\end{tabular}
\par\vspace{2pt}
\parbox{\linewidth}{\footnotesize Means over $n$ training seeds; largest standard deviation over seeds: 0.62\dB{} (FP32 twin, $60^\circ$). Macro: as in Table~\ref{tab:acc}; 40\dB: that SNR point; $60^\circ$ and zero-shot A--F: Sec.~\ref{sec:protocol}. 1-bit reference: phase-consistent proxy; FP32 twin: real weights and activations; no gate: blend with $g=0$, head fed the unscaled $\tilde b$; L2 proxy: symbol-average over symbol-difference power. Dense: real-valued network on the LS average, one 52-unit or two 104-unit hidden layers. KiB: packed (dense: FP32) parameters.\par}
\end{table}

\subsection{Bit-exact implementation}
\label{sec:verify}
Every integer-domain link from the exported weights to the board output is checked word for word.
The RTL and an independent integer model agree on 393,216 feature-map values, 1,536 channel-estimate components and 36 scalars over 12 frames, and the RTL is bit-exact against the C++ fixed-point model.
The floating-point checkpoint deviates from the integer model by $-34.6$\dB{} NMSE on average ($-27.4$\dB{} at the 99th percentile) over the 13,824 test pairs (macro NMSE: Sec.~\ref{sec:acc}).
Six board cases (three per MCS, ten packets each) return 1,280 CSI words each. Of the 7,680 words, 3,840 come from the early-LS window and 3,840 from the refined window (3,120 RG-BRN estimates, 240 edge tones and 480 DC/guard zeros).
All CSI payloads and sample indices match RTL, and all six complete record streams are byte-identical.
Back-to-back operation is bit-exact against the integer model in saturated 12- and 16-frame RTL runs, with per-cycle ownership assertions (Sec.~\ref{sec:schedule}).

\subsection{Packet error rate}
\label{sec:per}
\begin{figure*}[!t]
\centering
\includegraphics[width=\textwidth]{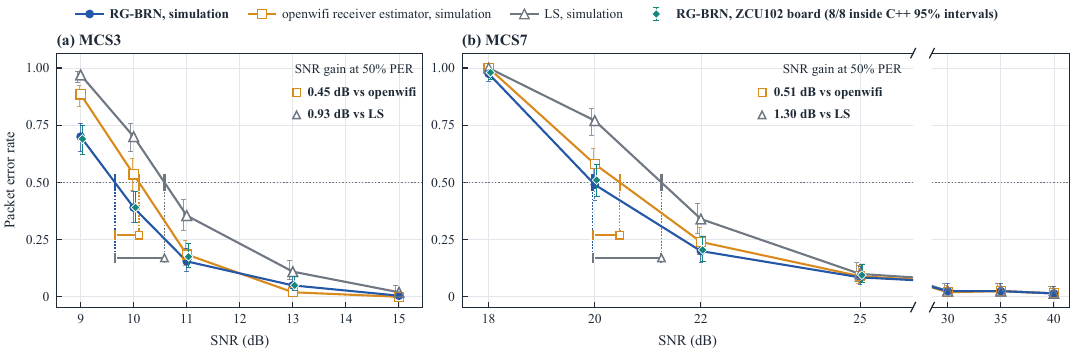}
\caption{PER on TGn-B (Wilson 95\% intervals) for (a)~MCS3 and (b)~MCS7: the three estimators in the RTL-simulated receiver (RG-BRN and LS as C++ models) and RG-BRN on the ZCU102 board. The SNR axis of (b) breaks after 26\dB{} and compresses 30--40\dB; otherwise both panels share one dB scale. At MCS7 from 30\dB{} the three PERs coincide within the intervals.}
\label{fig:per}
\end{figure*}

The NMSE gain translates into packets: where the openwifi receiver estimator's PER lies in $(0.1,0.9)$, RG-BRN removes 20.2\% [14.8, 25.6] of its packet errors and 38.3\% [33.7, 43.0] of LS's (Fig.~\ref{fig:per}).
\begin{figure*}[!t]
\centering
\includegraphics[width=\textwidth]{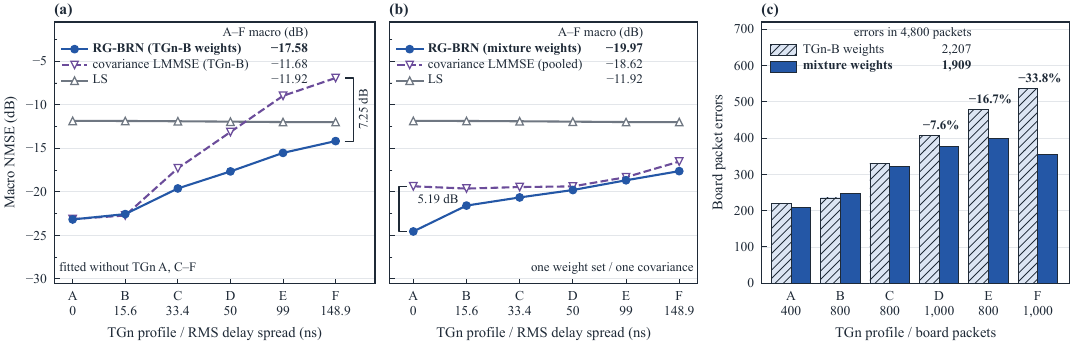}
\caption{TGn A--F. (a), (b)~Macro NMSE (2,304 pairs per profile) against profile and root-mean-square (RMS) delay spread (profile A is flat; brackets: largest lead of RG-BRN over the LMMSE). (a)~Fitted on TGn-B (RG-BRN: in its last training stage): RG-BRN with TGn-B weights and covariance LMMSE; LS as reference. (b)~Without profile information: RG-BRN with mixture weights, pooled-covariance LMMSE and LS. (c)~ZCU102 board packet errors per profile (packets below each), TGn-B against mixture weights on the same packets; percentages: error change on D--F.}
\label{fig:af}
\end{figure*}%
There RG-BRN makes 387 errors in 1,000 packets against 485 (openwifi) and 627 (LS); over all eight board-run points, 609 against 707 and 869.
Pooled over the five points, RG-BRN alone decodes 144 and the openwifi receiver estimator alone 46 of 1,000 paired packets; point by point, RG-BRN is significantly better than that estimator at MCS3 9 and 10\dB{} and MCS7 20\dB{} (exact McNemar, Holm-adjusted over ten tests, $p<0.02$) and than LS at all five points ($p<3\times10^{-7}$).
The SNR for 50\% PER falls by 0.45\dB{} [0.27, 0.61] and 0.51\dB{} [0.18, 0.76] (MCS3, MCS7) against the openwifi receiver estimator and by 0.93\dB{} [0.77, 1.09] and 1.30\dB{} [1.00, 1.51] against LS (brackets in Fig.~\ref{fig:per}).
All 160 overlapping packet outcomes match the integer RTL simulation. The 1,600-packet board PER lies within the C++ Wilson 95\% intervals at all eight points (619 errors against 609).

\begin{table}[!t]
\centering
\caption{Routed Implementation on the ZCU102}
\label{tab:impl}
\scriptsize
\setlength{\tabcolsep}{6.5pt}
\begin{tabular}{@{}lrrrr@{}}
\toprule
Scope & LUT & FF & DSP & RAMB18/36 \\
\midrule
RG-BRN accelerator & 179,189 & 172,363 & 1,622 & 1,080/0 \\
\;+ CDC adapter and FIFOs & 182,469 & 178,794 & 1,622 & 1,081/3 \\
Receiver + estimator & 202,416 & 202,901 & 1,731 & 1,086/10 \\
\midrule
\multicolumn{5}{@{}l}{\emph{Accelerator modules}} \\
\;Block engines (2) & 146,056 & 65,151 & 512 & 56/0 \\
\;Frontend & 13,314 & 12,677 & 343 & 0/0 \\
\;Backend & 18,800 & 17,284 & 767 & 0/0 \\
\;Boundary maps and control & 1,031 & 77,251 & 0 & 1,024/0 \\
\midrule
Estimator, \% of device & 66.6\% & 32.6\% & 64.4\% & 59.6\% \\
Receiver + estimator, \% of device & 73.9\% & 37.0\% & 68.7\% & 60.6\% \\
\midrule
\multicolumn{5}{@{}p{\dimexpr\columnwidth-2pt\relax}@{}}{Full design incl.\ DMA: WNS $+0.029$\,ns, WHS $+0.006$\,ns, 0 of 787,266 setup endpoints failing.} \\
\bottomrule
\end{tabular}
\par\vspace{2pt}
\parbox{\linewidth}{\footnotesize XCZU9EG, Vivado 2025.2, 200.02/100.01\,MHz. Estimator: accelerator plus adapter; block-RAM device share counts RAMB36 tiles; FF: flip-flops; WHS: worst hold slack; module LUTs sum to 12 more than the accelerator (cross-hierarchy LUT combining).\par}
\end{table}

\begin{table}[!t]
\centering
\caption{Design Space of the Channel-First Engine}
\label{tab:dse}
\scriptsize
\setlength{\tabcolsep}{3.0pt}
\begin{tabular}{@{}rcrrrrrrr@{}}
\toprule
PE/engine & Banks & Datapaths & Cycles & $\mu$s & LUT & FF & DSP & RAMB18 \\
\midrule
64 & 64/4 & 256 & 1,417 & 7.09$^\dagger$ & 185,093 & 176,072 & 1,622 & 1,080 \\
64 & 32/2 & 256 & 2,105 & 10.53$^\dagger$ & 191,203 & 189,413 & 1,622 & 540 \\
64 & 16/1 & 256 & 3,547 & 17.74\phantom{$^\dagger$} & 193,017 & 194,645 & 1,622 & 270 \\
32 & 64/4 & 128 & 1,865 & 9.33$^\dagger$ & 144,106 & 159,624 & 1,366 & 1,080 \\
32 & 32/2 & 128 & 2,119 & 10.60$^\dagger$ & \multicolumn{3}{c}{\emph{RTL latency only}} & 540 \\
32 & 16/1 & 128 & 3,561 & 17.81\phantom{$^\dagger$} & \multicolumn{3}{c}{\emph{RTL latency only}} & 270 \\
\bottomrule
\end{tabular}
\par\vspace{2pt}
\parbox{\linewidth}{\footnotesize Datapaths: PE/engine $\times$ 2 engines $\times$ 2 lanes. Cycles: isolated latency, 12-frame RTL simulation (5\,ns/cycle); $\dagger$: ready by the 17.08-$\mu$s bound (receive cycle: Cycles$/2{+}141$, derived). Resources: out-of-context post-synthesis, accelerator + frame buffers, 5-ns constraint, pre-route; RAMB18 of the latency-only rows follows the bank profile. Banks 8/1: 32.02\,$\mu$s (64+64) and 32.09\,$\mu$s (32+32).\par}
\end{table}

\subsection{Latency, cost and power}
\label{sec:hw}
The complete receiver with the estimator fits one XCZU9EG at 200/100\MHz (estimator: 66.6\% of the LUTs and 64.4\% of the DSP slices; receiver and estimator: 73.9\% and 68.7\%), and the full routed design closes timing at a worst negative slack (WNS) of $+0.029$\,ns (Table~\ref{tab:impl}).
In RTL simulation the estimate keeps an 8.59-$\mu$s margin to the 17.08-$\mu$s readiness bound, and at the 673-cycle saturated interval (Sec.~\ref{sec:schedule}) the accelerator delivers $2.97\times10^{5}$ estimates, or $3.3\times10^{12}$ binary products, per second at 200\MHz{} (derived).
Of the 11.13\,M convolution operations per estimate, 98.9\% are XNOR-popcount products in logic; the two block engines (146,056 LUTs, 81.5\% of the accelerator) use 512 DSP slices only for per-channel scales and slopes, two per PE datapath, while the real-valued frontend and backend use the other 1,110; the eight 24-bit boundary maps hold 94.8\% of the RAMB18.
The INA226 monitors~\cite{ti_ina226} measure an average of 2.67\,W on the complete design's PL rails, including DMA (684 readings over the 1,600-packet run at 0.25\% streaming duty). Idle power is 2.96\,W after configuration and reset release and 2.62\,W after the run.

\subsection{Design space}
\label{sec:dse}
One RTL spans a family of designs: PE count per engine (64 or 32) and bank profile (banks per map) set the isolated latency between 7.09 and 32.09\us, all eight configurations are bit-exact over 12 frames, and four (64/4 and 32/2 banks at both PE counts) meet the readiness bound with over 500 receive cycles to spare (derived; Table~\ref{tab:dse}).
At 64/4 banks, using 32 PEs per engine reduces the post-synthesis resource counts from 185,093 to 144,106 LUTs and from 1,622 to 1,366 DSP slices, with isolated latency increasing from 1,417 to 1,865 cycles.
The 32/2 bank profile halves map memory to 540 RAMB18 and needs 2,105 cycles with 64 PEs per engine and 2,119 cycles with 32 PEs per engine.

\subsection{Comparison with prior hardware}
\label{sec:compare}
Against published FPGA hardware (Table~\ref{tab:compare}), the 14 body convolutions of RG-BRN use 1-bit weights and activations and a 24-bit residual stream; the parameter set (61.34\,KiB packed, 11.6$\times$ below FP32) maps to integer constant ROMs in logic, where the tabulated learned estimators use 18- to 26-bit fixed point~\cite{haq2024lclsdnn,sharma2024lsidnn} or 8-bit weights~\cite{soltani2022receiver}.
To our knowledge, within the surveyed literature, no prior work reports a learned channel estimator inside a complete IEEE 802.11 receiver on an FPGA that combines (a) bit-exact agreement with its RTL by board read-back, (b) a board PER over replayed baseband of at least 1,000 packets and (c) power measured on the board with an INA226 monitor.

\begin{table*}[!t]
\centering
\caption{FPGA Channel-Estimation Hardware}
\label{tab:compare}
\scriptsize
\setlength{\tabcolsep}{2.4pt}
\providecommand{\rowcolor}[1]{}
\newcolumntype{Y}[1]{>{\raggedright\arraybackslash}p{#1}}
\begin{tabular}{Y{1.35cm}Y{2.95cm}Y{1.45cm}Y{2.9cm}Y{2.9cm}Y{2.55cm}Y{2.8cm}}
\toprule
Work & Estimator, precision & Device, clock & Latency (basis; budget) & Resources as printed (scope) & Power (method) & Accuracy protocol; evidence level \\
\midrule
\rowcolor{cBlue!10}
\textbf{This work} & RG-BRN; 1-bit body weights and activations, 24-bit maps; 179,544 parameters; 11.0\,M binary products + 0.12\,M conv.\ MACs & XCZU9EG, 200/100\MHz & 7.09\us{} core, isolated (S); ready 8.49\us{} (S, RTL system); budget: readiness bound 17.08\us & estimator + adapter, routed: 182,469 LUT, 178,794 FF, 1,622 DSP, 1,081 RAMB18 + 3 RAMB36 & 2.67\,W, mean over a 1,600-packet run, PL rails of the complete design incl.\ DMA (INA226, board) & NMSE vs LS, smoothers, LMMSE; board read-back = RTL; board PER, replayed: 1,600 (TGn-B), 4,800 per set (A--F)\\
\midrule
Havinga et al.~\cite{havinga2023hls} & classical CE + equalizer (openwifi) & Z-7020, 100\MHz & 1.62/1.66\us{} worst case, CE and smoothing (S); budget 3.6\us{} per symbol & CE + equalizer (HDL build): 5,417 LUT, 10,073 FF, 1 BRAM, 19 DSP & -- & PER against a reference; board PER, 10,000 HT packets \\
\addlinespace[2pt]
Haq et al.~\cite{haq2024lclsdnn} & LS + DNN (802.11p); fixed $\langle24,8\rangle$/$\langle18,2\rangle$ & ZC706, 200\MHz & 2.52\us{}, estimator execution (B) & estimator: 86,165 LUT, 89,244 FF, 9 BRAM, 172 DSP & 5.76\,W total, PL 4.23\,W & NMSE, BER curves \\
\addlinespace[2pt]
Sharma et al.~\cite{sharma2024lsidnn} & LS + DL (3GPP comb-pilot, three channel models); fixed (26,8) & ZC706 & -- & estimator: 2,554 LUT, 3,021 FF, 8 BRAM, 6 DSP & 2.05\,W total & -- \\
\addlinespace[2pt]
Soltani et al.~\cite{soltani2022receiver} & MLP on the 802.11a L-LTF; 8-bit weights & Kintex-7 T410 (X310), 100\MHz & 0.33\,ms, compressed MLP (B) & -- & -- & BER; over-the-air BER; latency measured on the X310 \\
\bottomrule
\end{tabular}
\par\vspace{2pt}
\parbox{\linewidth}{\footnotesize Prior work as printed by its authors (--: not available); latency basis S: simulation, B: board measurement; resources carry their scope. CE: channel estimation; DNN: deep neural network; DL: deep learning; HDL: hardware description language; MLP: multilayer perceptron; BER: bit error rate; BRAM: block RAM; MAC: multiply-accumulate.\par}
\end{table*}

\subsection{TGn A--F delay profiles}
\label{sec:af}
One mixture-trained weight set (last training stage on 54,000 A--F pairs) covers the delay-spread range of TGn A--F without profile information: it leads a pooled-covariance LMMSE (one covariance fitted on the same A--F mixture) in macro NMSE on every profile, $-19.97$ against $-18.62$\dB{} on A--F (1.35\dB, interval 1.29--1.40\dB).
The lead is 0.36\dB{} (E) to 5.19\dB{} (A) per profile (Fig.~\ref{fig:af}b) and 0.95\dB{} on B--F; the A--F lead comes from 0--15\dB{} (0.84--1.53\dB{} per SNR; first of three final-stage seeds, A--F macro standard deviation $\le$0.05\dB).
When both are fitted on TGn-B (the covariance for the LMMSE, the last of three training stages for RG-BRN), RG-BRN leads by 5.90\dB{} on A--F (Fig.~\ref{fig:af}a): it is ahead by 2.32, 4.53, 6.55 and 7.25\dB{} on \mbox{C--F} and level on A, while the LMMSE's NMSE is 0.16\dB{} lower on B, its fitting profile.
From TGn-B to F, RG-BRN's NMSE rises by 8.4\dB, about half the TGn-B-fitted LMMSE's 15.8\dB, and RG-BRN stays ahead of LS in macro NMSE on all six profiles.
Trained in one TGn-B stage only, RG-BRN still leads that LMMSE by 4.1\dB{} on A--F.

A new weight set is a re-synthesis and re-implementation of the same RTL and schedule (15,227 of the 172,032 binary weights, 8.85\%, change sign); both weight-set builds meet 200/100\MHz{} timing, and CSI read-back of the mixture build (three cases, 3,840 words) equals its RTL receiver's.
On the board, the mixture weights make 1,909 errors against 2,207 for the TGn-B weights in 4,800 packets each (13.5\% fewer, 95\% channel-bootstrap interval [11.4, 15.6]; 418 packets decoded only by the mixture weights, 120 only by the TGn-B weights; Fig.~\ref{fig:af}c). The 4,800 packets are 200 further packets at each of 24 points picked from a 40-packet screening by a pre-fixed rule: per profile and MCS, two or three points with both PERs in $[0.08,0.92]$, else the pair crossing mean PER 0.5 (none for A at MCS7).
The largest error reductions occur on D--F: 33.8\% fewer errors on F, 16.7\% on E and 7.6\% on D, with all eight significant points (exact McNemar, Holm-adjusted $p<0.005$) on E and F; profiles A--C stay within 5.1\% with no significant point.

\section{Conclusion}
\label{sec:conclusion}
RG-BRN raises estimate accuracy where links are weakest.
A closed-form gate from the two L-LTF symbols adapts a network whose 14 binary convolutions receive no SNR input: 61.34\,KiB of parameters reach $-22.57$\dB{} macro NMSE on TGn-B, 6.2\dB{} below the openwifi smoother and within 0.2\dB{} of covariance-fitted LMMSE. Where the openwifi receiver estimator's PER is 10--90\%, RG-BRN removes 20.2\% of its packet errors in simulation.
In single-stage retrainings, the gate gains 5.1\dB{} over an ungated network at 40\dB{} on TGn-B, and the 1-bit-body network leads its FP32 twin by 2.2\dB{} under a $60^\circ$ common phase on the second L-LTF symbol.
One mixture-trained weight set leads pooled-covariance LMMSE in macro NMSE on every TGn profile \mbox{A--F} (1.35\dB{} in \mbox{A--F} macro), and on the board it removes 13.5\% of the \mbox{TGn-B} weights' packet errors at 24 A--F transition points.
The channel-first engine combines paired even/odd-bin transactions, unfolded PEs and an overlapped frontend to achieve a 7.09-$\mu$s isolated latency.
On a ZCU102 board with replayed baseband, 7,680 CSI words equal the RTL receiver's, the 1,600-packet PER lies inside software 95\% intervals and the complete design averages 2.67\,W on its PL rails. In RTL system simulation, a \mbox{1-bit}-body learned estimator in a complete 802.11 receiver is ready 8.59\us{} before the readiness bound.

\bibliographystyle{IEEEtran}
\IEEEtriggeratref{34}
\bibliography{refs}

@standard{ieee2025std80211,
  title = {{IEEE} Standard for Information Technology--Telecommunications and Information Exchange between Systems Local and Metropolitan Area Networks--Specific Requirements Part 11: Wireless {LAN} Medium Access Control ({MAC}) and Physical Layer ({PHY}) Specifications},
  organization = {IEEE},
  number = {802.11-2024},
  year = {2025},
  doi = {10.1109/IEEESTD.2025.10979691}
}

@techreport{erceg2004tgn,
  author = {V. Erceg and L. Schumacher and P. Kyritsi and A. Molisch and D. S. Baum and A. Y. Gorokhov and others},
  title = {{TGn} Channel Models},
  institution = {IEEE P802.11 Wireless LANs},
  type = {IEEE document},
  number = {IEEE 802.11-03/940r4},
  year = {2004},
  month = may
}

@inproceedings{jiao2020openwifi,
  author = {Xianjun Jiao and Wei Liu and Michael Mehari and Muhammad Aslam and Ingrid Moerman},
  title = {{openwifi}: a free and open-source {IEEE802.11 SDR} implementation on {SoC}},
  booktitle = {2020 IEEE 91st Vehicular Technology Conference (VTC2020-Spring)},
  pages = {1--2},
  year = {2020},
  doi = {10.1109/vtc2020-spring48590.2020.9128614}
}

@misc{shi2022openofdm,
  author = {Shi, Jinghao},
  title = {{OpenOFDM}: Synthesizable, Modular {Verilog} Implementation of 802.11 {OFDM} Decoder},
  howpublished = {\url{https://github.com/jhshi/openofdm}; documentation: \url{https://openofdm.readthedocs.io/en/latest/}},
  note = {Accessed 2026-11-25}
}

@misc{openwifihw,
  author = {Jiao, Xianjun and others},
  title = {{openwifi-hw}: Hardware ({FPGA}) Design of the {openwifi} {IEEE} 802.11 Transceiver},
  howpublished = {\url{https://github.com/open-sdr/openwifi-hw}},
  note = {Accessed 2026-11-25}
}

@article{edfors1998svd,
  author = {O. Edfors and M. Sandell and J.-J. van de Beek and S. K. Wilson and P. O. B{\"o}rjesson},
  title = {{OFDM} channel estimation by singular value decomposition},
  journal = {IEEE Trans. Commun.},
  volume = {46},
  number = {7},
  pages = {931--939},
  year = {1998},
  doi = {10.1109/26.701321}
}

@article{ye2018power,
  author = {Ye, Hao and Li, Geoffrey Ye and Juang, Biing-Hwang},
  title = {Power of Deep Learning for Channel Estimation and Signal Detection in {OFDM} Systems},
  journal = {IEEE Wireless Commun. Lett.},
  volume = {7},
  number = {1},
  pages = {114--117},
  year = {2018},
  doi = {10.1109/lwc.2017.2757490}
}

@inproceedings{guler2025adafortitran,
  author = {Guler, Berkay and Jafarkhani, Hamid},
  title = {{AdaFortiTran}: An Adaptive Transformer Model for Robust {OFDM} Channel Estimation},
  booktitle = {ICC 2025 - IEEE International Conference on Communications},
  pages = {3797--3802},
  year = {2025},
  doi = {10.1109/icc52391.2025.11160810}
}

@article{haq2023lsdnn,
  author = {Syed Asrar Ul Haq and Abdul Karim Gizzini and Shakti Shrey and Sumit J. Darak and Sneh Saurabh and Marwa Chafii},
  title = {Deep Neural Network Augmented Wireless Channel Estimation for Preamble-Based {OFDM} {PHY} on {Zynq} System on Chip},
  journal = {IEEE Trans. Very Large Scale Integr. (VLSI) Syst.},
  volume = {31},
  number = {7},
  pages = {1026--1038},
  year = {2023},
  doi = {10.1109/tvlsi.2023.3274555}
}

@inproceedings{haq2024lclsdnn,
  author = {Haq, Syed Asrar Ul and Singh, Varun and Tanaji, Bhanu Teja and Darak, Sumit},
  title = {Low Complexity High Speed Deep Neural Network Augmented Wireless Channel Estimation},
  booktitle = {2024 37th International Conference on VLSI Design and 2024 23rd International Conference on Embedded Systems (VLSID)},
  pages = {235--240},
  year = {2024},
  doi = {10.1109/vlsid60093.2024.00045}
}

@article{sharma2024lsidnn,
  author = {Animesh Sharma and Syed Asrar Ul Haq and Sumit J. Darak},
  title = {Low Complexity Deep Learning Augmented Wireless Channel Estimation for Pilot-Based {OFDM} on {Zynq} System on Chip},
  journal = {IEEE Trans. Circuits Syst. I, Reg. Papers},
  volume = {71},
  number = {5},
  pages = {2334--2347},
  year = {2024},
  doi = {10.1109/TCSI.2024.3371780}
}

@inproceedings{hubara2016bnn,
  author = {Hubara, Itay and Courbariaux, Matthieu and Soudry, Daniel and El-Yaniv, Ran and Bengio, Yoshua},
  title = {Binarized Neural Networks},
  booktitle = {Advances in Neural Information Processing Systems 29 ({NIPS} 2016)},
  pages = {4107--4115},
  year = {2016},
  url = {https://papers.nips.cc/paper_files/paper/2016/hash/d8330f857a17c53d217014ee776bfd50-Abstract.html}
}

@inproceedings{umuroglu2017finn,
  author = {Umuroglu, Yaman and Fraser, Nicholas J. and Gambardella, Giulio and Blott, Michaela and Leong, Philip and Jahre, Magnus and Vissers, Kees},
  title = {{FINN}: A Framework for Fast, Scalable Binarized Neural Network Inference},
  booktitle = {Proc. ACM/SIGDA Int. Symp. Field-Programmable Gate Arrays (FPGA)},
  pages = {65--74},
  year = {2017},
  doi = {10.1145/3020078.3021744},
}

@article{lu2021bcsinet,
  author = {Lu, Zhilin and Wang, Jintao and Song, Jian},
  title = {Binary Neural Network Aided {CSI} Feedback in Massive {MIMO} System},
  journal = {IEEE Wireless Commun. Lett.},
  volume = {10},
  number = {6},
  pages = {1305--1308},
  year = {2021},
  doi = {10.1109/lwc.2021.3064963}
}

@article{lu2022acrnet,
  author = {Lu, Zhilin and Zhang, Xudong and He, Hongyi and Wang, Jintao and Song, Jian},
  title = {Binarized Aggregated Network With Quantization: Flexible Deep Learning Deployment for {CSI} Feedback in Massive {MIMO} Systems},
  journal = {IEEE Trans. Wireless Commun.},
  volume = {21},
  number = {7},
  pages = {5514--5525},
  year = {2022},
  doi = {10.1109/TWC.2022.3141653},
}

@article{shankar2024binarized,
  author = {Shankar, Nitin Priyadarshini and Sadhukhan, Deepsayan and Nayak, Nancy and Tholeti, Thulasi and Kalyani, Sheetal},
  title = {Binarized {ResNet}: Enabling Robust Automatic Modulation Classification at the Resource-Constrained Edge},
  journal = {IEEE Trans. Cogn. Commun. Netw.},
  volume = {10},
  number = {5},
  pages = {1913--1927},
  year = {2024},
  doi = {10.1109/TCCN.2024.3391325},
}

@inproceedings{boumard2003snr,
  author = {Boumard, Sandrine},
  title = {Novel noise variance and {SNR} estimation algorithm for wireless {MIMO} {OFDM} systems},
  booktitle = {GLOBECOM '03. IEEE Global Telecommunications Conference},
  address = {San Francisco, CA, USA},
  volume = {3},
  pages = {1330--1334},
  month = dec,
  year = {2003},
  doi = {10.1109/GLOCOM.2003.1258454}
}

@inproceedings{zivkovic2009preamble,
  author = {Zivkovic, Milan and Mathar, Rudolf},
  title = {Preamble-Based {SNR} Estimation in Frequency Selective Channels for Wireless {OFDM} Systems},
  booktitle = {VTC Spring 2009 - IEEE 69th Vehicular Technology Conference},
  year = {2009},
  doi = {10.1109/vetecs.2009.5073813}
}

@article{soltani2022receiver,
  author = {Soltani, Nasim and Cheng, Hai and Belgiovine, Mauro and Li, Yanyu and Li, Haoqing and Azari, Bahar and D'Oro, Salvatore and Imbiriba, Tales and Melodia, Tommaso and Closas, Pau and Wang, Yanzhi and Erdogmus, Deniz and Chowdhury, Kaushik},
  title = {Neural Network-Based {OFDM} Receiver for Resource Constrained {IoT} Devices},
  journal = {IEEE Internet Things Mag.},
  volume = {5},
  number = {3},
  pages = {158--164},
  year = {2022},
  doi = {10.1109/iotm.001.2200051}
}

@article{brennsteiner2022receiver,
  author = {Brennsteiner, Stefan and Arslan, Tughrul and Thompson, John and McCormick, Andrew},
  title = {A Real-Time Deep Learning {OFDM} Receiver},
  journal = {ACM Trans. Reconfigurable Technol. Syst.},
  volume = {15},
  number = {3},
  pages = {1--25},
  year = {2022},
  doi = {10.1145/3494049}
}

@article{lyu2026swiftchannel,
  author = {Lyu, Shengzhe and She, Yuhan and Duan, Di and Ni, Tao and Chan, Yu Hin and Luo, Chengwen and Cheung, Ray C. C. and Xu, Weitao},
  title = {{SwiftChannel}: Algorithm-Hardware Co-Design for Deep Learning-Based {5G} Channel Estimation},
  journal = {IEEE Trans. Mobile Comput.},
  volume = {25},
  number = {6},
  pages = {8268--8284},
  month = jun,
  year = {2026},
  doi = {10.1109/TMC.2025.3647644}
}

@article{guo2025riscvce,
  author = {Meng Guo and Qi Wu and Chuanning Wang and Yangcan Zhou and Shaowei Wang and Chuan Zhang and Zhongfeng Wang and Jun Lin},
  title = {A {RISC-V} Domain-Specific Processor for Deep Learning-Based Channel Estimation},
  journal = {IEEE Trans. Circuits Syst. I, Reg. Papers},
  volume = {72},
  number = {5},
  pages = {2380--2393},
  year = {2025},
  doi = {10.1109/TCSI.2025.3547319}
}

@article{yun2026vitofdm,
  author = {Sangbu Yun and Chanhee Lee and Soonhyun Kwon and Jeongtaek Chang and Zhengya Zhang and Youngjoo Lee},
  title = {A {14.4-Gb/s} {7.3-pJ/b} Universal Vision Transformer {OFDM} Channel Estimation Processor for {B5G/6G} Communication Systems},
  journal = {IEEE Trans. Circuits Syst. I, Reg. Papers},
  year = {2026},
  doi = {10.1109/TCSI.2026.3729434},
  note = {early access}
}

@article{chundi2022fpgace,
  author = {Chundi, Pavan Kumar and Wang, Xiaodong and Seok, Mingoo},
  title = {Channel Estimation Using Deep Learning on an {FPGA} for {5G} Millimeter-Wave Communication Systems},
  journal = {IEEE Trans. Circuits Syst. I, Reg. Papers},
  volume = {69},
  number = {2},
  pages = {908--918},
  year = {2022},
  doi = {10.1109/TCSI.2021.3117886}
}

@article{li2023ndrnet,
  author = {Li, Yinying and Bian, Xin and Li, Mingqi},
  title = {Denoising Generalization Performance of Channel Estimation in Multipath Time-Varying {OFDM} Systems},
  journal = {Sensors},
  volume = {23},
  number = {6},
  pages = {3102},
  year = {2023},
  doi = {10.3390/s23063102}
}

@article{elazizi2026smartce,
  author = {El-azizi, Ousama and Fateh, Rachid and Safi, Said},
  title = {A Spatio-temporal Residual Attention Network with {SNR}-adaptive Weighting for {MIMO-OFDM} Channel Estimation},
  journal = {J. Telecommun. Inf. Technol.},
  volume = {105},
  number = {3},
  pages = {52--64},
  year = {2026},
  doi = {10.26636/jtit.2026.3.2688}
}

@inproceedings{zhou2026gsldiffusion,
  author = {Zhou, Jixing and Huang, Xinming},
  title = {{SNR}-Gated {LSTM}-Conditioned Diffusion Model for {MIMO} Channel Estimation},
  booktitle = {2026 IEEE 104th Vehicular Technology Conference (VTC2026-Fall)},
  pages = {1--5},
  month = sep,
  year = {2026},
  doi = {10.1109/VTC2026-Fall68024.2026.11758496},
  note = {arXiv:2610.08977}
}

@inproceedings{vanlier2020optcomnet,
  author = {van Lier, Michel and Balatsoukas-Stimming, Alexios and Corporaal, Henk and Zivkovic, Zoran},
  title = {{OPTCOMNET}: Optimized Neural Networks for Low-Complexity Channel Estimation},
  booktitle = {ICC 2020 - 2020 IEEE International Conference on Communications (ICC)},
  pages = {1--6},
  year = {2020},
  doi = {10.1109/icc40277.2020.9149049}
}

@article{gan2022sasseinet,
  author = {Gan, Jiayan and Hu, Ang and Kang, Ziyi and Qu, Zhipeng and Yang, Zhanxiang and Yang, Rui and Wang, Yibing and Shao, Huaizong and Zhou, Jun},
  title = {{SAS-SEINet}: A {SNR}-Aware Adaptive Scalable {SEI} Neural Network Accelerator Using Algorithm--Hardware Co-Design for High-Accuracy and Power-Efficient {UAV} Surveillance},
  journal = {Sensors},
  volume = {22},
  number = {17},
  pages = {6532},
  year = {2022},
  doi = {10.3390/s22176532}
}

@article{gao2018comnet,
  author = {Gao, Xuanxuan and Jin, Shi and Wen, Chao-Kai and Li, Geoffrey Ye},
  title = {{ComNet}: Combination of Deep Learning and Expert Knowledge in {OFDM} Receivers},
  journal = {IEEE Commun. Lett.},
  volume = {22},
  number = {12},
  pages = {2627--2630},
  year = {2018},
  doi = {10.1109/lcomm.2018.2877965}
}

@patent{wallace2012noise,
  author = {Wallace, Mark S. and Monsen, Peter},
  title = {Noise estimation for wireless communication},
  nationality = {U.S.},
  number = {8\,107\,566},
  day = {31},
  month = jan,
  year = {2012}
}

@inproceedings{xia2023bbcu,
  author = {Xia, Bin and Zhang, Yulun and Wang, Yitong and Tian, Yapeng and Yang, Wenming and Timofte, Radu and Van Gool, Luc},
  title = {Basic Binary Convolution Unit for Binarized Image Restoration Network},
  booktitle = {Proc. Int. Conf. Learning Representations (ICLR)},
  year = {2023},
  note = {arXiv:2210.00405},
}

@inproceedings{tridgell2019amc,
  author = {Tridgell, Stephen and Boland, David and Leong, Philip H. W. and {Siddhartha}},
  title = {Real-Time Automatic Modulation Classification},
  booktitle = {2019 International Conference on Field-Programmable Technology (ICFPT)},
  pages = {299--302},
  year = {2019},
  doi = {10.1109/icfpt47387.2019.00052}
}

@inproceedings{defreitas2026fpgaeq,
  author = {de Freitas, Icaro Menezes Saraiva and Soares, Jonathan Aguiar and Le{\~a}o, Pedro Arthur Bessa and Rothenberg, Christian Esteve and Arantes, Dalton S.},
  title = {{FPGA}-Embedded Machine Learning Equalizer for a {MIMO} {BPSK} System on a {ZCU104} {MPSoC}},
  booktitle = {XLIV Simp{\'o}sio Brasileiro de Telecomunica{\c{c}}{\~o}es e Processamento de Sinais (SBrT)},
  year = {2026},
  doi = {10.14209/sbrt.2026.1571300000}
}

@article{gupta2025neuromorphicrx,
  author = {Gupta, Ankit and Dizdar, Onur and Chen, Yun and Kadan, Fehmi Emre and Sattarzadeh, Ata and Wang, Stephen},
  title = {{NeuromorphicRx}: From Neural to Spiking Receiver},
  journal = {IEEE Trans. Wireless Commun.},
  volume = {25},
  pages = {10161--10174},
  year = {2026},
  doi = {10.1109/TWC.2025.3650300}
}

@article{carreras2020tcn,
  author = {Carreras, Marco and Deriu, Gianfranco and Raffo, Luigi and Benini, Luca and Meloni, Paolo},
  title = {Optimizing Temporal Convolutional Network Inference on {FPGA}-Based Accelerators},
  journal = {IEEE J. Emerg. Sel. Topics Circuits Syst.},
  volume = {10},
  number = {3},
  pages = {348--361},
  month = sep,
  year = {2020},
  doi = {10.1109/JETCAS.2020.3014503}
}

@article{wu2024dilatedfpga,
  author = {Tsung-Hsi Wu and Chang Shu and Tsung-Te Liu},
  title = {An Efficient {FPGA}-Based Dilated and Transposed Convolutional Neural Network Accelerator},
  journal = {IEEE Trans. Circuits Syst. I, Reg. Papers},
  volume = {71},
  number = {11},
  pages = {5178--5186},
  year = {2024},
  doi = {10.1109/TCSI.2024.3428636}
}

@inproceedings{zhou2021implicit,
  author = {Zhou, Yangjie and Yang, Mengtian and Guo, Cong and Leng, Jingwen and Liang, Yun and Chen, Quan and Guo, Minyi and Zhu, Yuhao},
  title = {Characterizing and Demystifying the Implicit Convolution Algorithm on Commercial Matrix-Multiplication Accelerators},
  booktitle = {2021 IEEE International Symposium on Workload Characterization (IISWC)},
  pages = {214--225},
  year = {2021},
  doi = {10.1109/IISWC53511.2021.00029}
}

@inproceedings{shen2017multiclp,
  author = {Shen, Yongming and Ferdman, Michael and Milder, Peter},
  title = {Maximizing {CNN} Accelerator Efficiency Through Resource Partitioning},
  booktitle = {Proceedings of the 44th Annual International Symposium on Computer Architecture},
  pages = {535--547},
  year = {2017},
  doi = {10.1145/3079856.3080221}
}

@inproceedings{zhao2021dualopu,
  author = {Zhao, Tiandong and Yu, Yunxuan and Wang, Kun and He, Lei},
  title = {Heterogeneous Dual-Core Overlay Processor for Light-Weight {CNNs}},
  booktitle = {2021 IEEE 29th Annual International Symposium on Field-Programmable Custom Computing Machines (FCCM)},
  year = {2021},
  pages = {264},
  doi = {10.1109/FCCM51124.2021.00051},
  note = {Extended text: arXiv:2110.01103}
}

@article{denblanken2025chameleon,
  author = {den Blanken, Douwe and Frenkel, Charlotte},
  title = {Chameleon: A Multiplier-Free Temporal Convolutional Network Accelerator for End-to-End Few-Shot and Continual Learning from Sequential Data},
  journal = {IEEE J. Solid-State Circuits},
  volume = {61},
  number = {7},
  pages = {3814--3829},
  month = jul,
  year = {2026},
  doi = {10.1109/JSSC.2025.3645640}
}

@inproceedings{havinga2023hls,
  author = {Havinga, Thijs and Jiao, Xianjun and Liu, Wei and Moerman, Ingrid},
  title = {Accelerating {FPGA}-Based {Wi-Fi} Transceiver Design and Prototyping by High-Level Synthesis},
  booktitle = {2023 IEEE 31st Annual International Symposium on Field-Programmable Custom Computing Machines (FCCM)},
  pages = {219},
  month = may,
  year = {2023},
  doi = {10.1109/FCCM57271.2023.00047},
  note = {Extended version: arXiv:2305.13351}
}

@article{hu2026bapetransformer,
  author = {Xianghong Hu and Wenjun Zhuo and Qihang Ye and Yuanmiao Lin and Xueming Li and Hongmin Huang and Ruidian Zhan and Shuting Cai and Xiaoming Xiong},
  title = {{BAPE}: A Binary or Ternary-Weights Transformer Accelerator Based on Precomputation at the Edge},
  journal = {IEEE Trans. Circuits Syst. I, Reg. Papers},
  volume = {73},
  number = {11},
  pages = {7254--7267},
  year = {2026},
  doi = {10.1109/TCSI.2026.3727987}
}

@manual{ti_ina226,
  author = {{Texas Instruments}},
  title = {{INA226} 36V, 16-Bit, Ultra-Precise {I2C} Output Current, Voltage, and Power Monitor With Alert},
  number = {SBOS547C},
  year = {2026},
  note = {{SBOS547C}}
}

@inproceedings{gizzini2020trfi,
  author = {Gizzini, Abdul Karim and Chafii, Marwa and Nimr, Ahmad and Fettweis, Gerhard},
  title = {Joint {TRFI} and Deep Learning for Vehicular Channel Estimation},
  booktitle = {2020 IEEE Globecom Workshops (GC Wkshps)},
  address = {Taipei, Taiwan},
  pages = {1--6},
  month = dec,
  year = {2020},
  doi = {10.1109/GCWkshps50303.2020.9367412}
}

@inproceedings{liu2020reactnet,
  author = {Liu, Zechun and Shen, Zhiqiang and Savvides, Marios and Cheng, Kwang-Ting},
  title = {{ReActNet}: Towards Precise Binary Neural Network with Generalized Activation Functions},
  booktitle = {Proc. European Conf. Computer Vision (ECCV)},
  pages = {143--159},
  year = {2020},
  doi = {10.1007/978-3-030-58568-6_9},
}

@inproceedings{liu2018bireal,
  author = {Liu, Zechun and Wu, Baoyuan and Luo, Wenhan and Yang, Xin and Liu, Wei and Cheng, Kwang-Ting},
  title = {{Bi-Real Net}: Enhancing the Performance of 1-Bit {CNNs} with Improved Representational Capability and Advanced Training Algorithm},
  booktitle = {Proc. European Conf. Computer Vision (ECCV)},
  pages = {747--763},
  year = {2018},
  doi = {10.1007/978-3-030-01267-0_44},
}

@article{wilson1927probable,
  author = {Edwin B. Wilson},
  title = {Probable Inference, the Law of Succession, and Statistical Inference},
  journal = {J. Amer. Statist. Assoc.},
  volume = {22},
  number = {158},
  pages = {209--212},
  year = {1927},
  doi = {10.1080/01621459.1927.10502953}
}

\end{document}